\documentclass[aps, prb, superscriptaddress, citeautoscript, showpacs, twocolumn]{revtex4-1}
\usepackage{float}
\usepackage[pdftex]{graphicx}
\usepackage{epstopdf}
\usepackage{amsmath}
\usepackage{subfigure}
\usepackage{fixltx2e}
\usepackage{dcolumn}

\begin{document}

\title{Correlating magnetic structure and magnetotransport \protect\\ in semimetal thin films of Eu$_{1-x}$Sm$_x$TiO$_3$}

\author{Zach Porter}
\thanks{These two authors contributed equally}
\affiliation{Materials Department, University of California, Santa Barbara, California 93106, USA}

\author{Ryan F. Need}
\thanks{These two authors contributed equally}
\affiliation{Department of Materials Science and Engineering, University of Florida, Gainesville, Florida 32611, USA}
\affiliation{NIST Center for Neutron Research, National Institute of Standards and Technology, Gaithersburg, Maryland 20899, USA}

\author{Kaveh Ahadi}
\affiliation{Materials Department, University of California, Santa Barbara, California 93106, USA}
\affiliation{Department of Materials Science and Engineering, North Carolina State University, Raleigh, North Carolina 27695, USA}

\author{Yang Zhao}
\author{Zhijun Xu}
\affiliation{NIST Center for Neutron Research, National Institute of Standards and Technology, Gaithersburg, Maryland 20899, USA}
\affiliation{Department of Materials Science and Engineering, University of Maryland, College Park, Maryland 20742, USA}

\author{Brian J. Kirby}
\author{Jeffrey W. Lynn}
\affiliation{NIST Center for Neutron Research, National Institute of Standards and Technology, Gaithersburg, Maryland 20899, USA}

\author{Susanne Stemmer}
\author{Stephen D. Wilson}
\affiliation{Materials Department, University of California, Santa Barbara, California 93106, USA}
\email[email: ]{stephendwilson@ucsb.edu}

\date{\today}

\begin{abstract}
We report on the evolution of the average and depth-dependent magnetic order in thin film samples of biaxially stressed and electron-doped EuTiO$_3$ for samples across a doping range $<$0.1 to 7.8 $\times 10^{20}$ cm$^{-3}$. Under an applied in-plane magnetic field, the G-type antiferromagnetic ground state undergoes a continuous spin-flop phase transition into in-plane, field-polarized ferromagnetism. The critical field for ferromagnetism slightly decreases with an increasing number of free carriers, yet the field evolution of the spin-flop transition is qualitatively similar across the doping range. Unexpectedly, we observe interfacial ferromagnetism with saturated Eu$^{2+}$ moments at the substrate interface at low fields preceding ferromagnetic saturation throughout the bulk of the degenerate semiconductor film. We discuss the implications of these findings for the unusual magnetotransport properties of this compound. 
\end{abstract}

\maketitle
\section{Introduction}

Weyl semimetals (WSMs) are a broadly sought class of topologically nontrivial semimetals \cite{yan2017topological}. Their protected, nondegenerate pairs of band crossings act as chiral Weyl points, which serve as sources and sinks of Berry flux in momentum space. The resulting high mobility and inherent spin-momentum locking of WSMs makes these materials strong candidates for a multitude of applications, spanning from spintronics \cite{Smejkal2018} to catalysis \cite{Rajamathi2017}. A WSM state requires breaking either inversion or time reversal symmetry in a degenerately doped semiconductor with one or more pairs of band crossings \cite{yan2017topological}. As a result, identifying candidate WSM states requires careful analysis of both atomic lattice and magnetic symmetries, and several examples of noncentrosymmetric \cite{huang2015weyl,lv2015experimental} and magnetically ordered \cite{li2019intrinsic,murakami2019realization,zhang2019topological} WSM systems have been uncovered. This study aims to explore relations between the Weyl physics and the crystallographic details of the magnetic order in one magnetic WSM candidate. 

Recent evidence suggests the presence of band crossings  and a WSM phase in Eu$_{1-x}$Sm$_x$TiO$_3$ films grown on (001)-oriented (La$_{0.3}$Sr$_{0.7})($Al$_{0.65}$Ta$_{0.35}$)O$_3$ (LSAT). While the parent compound EuTiO$_3$ is a band insulator due to its $3d^0$ Ti$^{4+}$ electronic configuration, small levels of Sm substitution introduce electrons into the primarily Ti $3d$ $t_{2g}$ conduction bands, driving a filling-controlled metal-insulator transition \cite{Ahadi2017}. At low electron densities, the measured anomalous Hall effect (AHE) resistance is negative. However, upon doping beyond $n=4{-}5 \times 10^{20}$ cm$^{-3}$ the AHE signal becomes positive, similar to behavior in La-substituted EuTiO$_3$ films \cite{Takahashi2009}. This AHE sign change is suggestive of a shifting in the chemical potential across one or more nodal band crossings \cite{Fang2003}. In the case of Eu$_{1-x}$Sm$_x$TiO$_3$, density functional theory (DFT) electronic structure calculations corroborate the presence of such a pair of band crossings along the $\Gamma$-X direction, in agreement with the details of the magnetotransport measurements \cite{Ahadi2018}. Taken together, the AHE and DFT results suggest a WSM phase.

In addition to the Weyl physics, anisotropic magnetoresistance (AMR) measurements of this material system reveal symmetry changes suggestive of field-driven magnetic phase transitions \cite{Ahadi2019}. The origins of these various transitions seen in AMR are however currently unclear, as are the underpinnings of other anomalous transverse magnetotransport in this material.  For instance, the field dependence of the AHE resistance is nonlinear, nonmonotonic, and \emph{not} proportional to magnetization \cite{Ahadi2018}. This is suggestive of a topological Hall effect (THE) originating in real or momentum space and evolves under applied field and doping \cite{Neubauer2009}, although alternative explanations have also been proposed \cite{Takahashi2018, Kan2018}. While these magnetotransport results clearly demonstrate strong coupling of Eu moments to the charge transport, the ability to interpret the mechanisms driving this behavior is limited by little direct knowledge of the underlying magnetic order in films. 

For context, we briefly digress to summarize the structure and magnetism in bulk samples. In the bulk, EuTiO$_3$ has a cubic perovskite structure at high temperatures and undergoes an antiferrodistortive transition due to antiphase tilting of oxygen along $c$ (as in SrTiO$_3$) at 280 K, yielding a tetragonal structure $I4/mcm$ with $a{=}b$ parameters 0.25$\%$ smaller than $c$ below 50 K \cite{Goian2012,Bessas2013}. Below T\textsubscript{N} = 6 K, EuTiO$_3$ orders as a G-type antiferromagnet (AFM) in which the $4f^7$ Eu$^{2+}$ sites have localized $S{=}7/2$ moments oriented in the $ab$ plane \cite{Scagnoli2012}. From the ordered zero-field state, application of an external magnetic field causes a spin-flop transition near $\mu_0 H=0.3$ T, though the behavior is strongly anisotropic \cite{Petrovic2013}. As the applied field approaches 1 T, the magnetization saturates near the expected 7 $\mu_B/$Eu in a roughly isotropic manner \cite{Petrovic2013}.

In EuTiO$_3$/LSAT(001) films, G-type AFM order was first observed via resonant X-ray magnetic scattering with a similar ordering temperature as bulk samples \cite{Ryan2013}. In this geometry, the substrate applies a 1$\%$ compressive epitaxial strain in the $ab$ plane and lengthens the unit cell along $c$, an effect confirmed by both X-ray diffraction and first-principles studies \cite{Ryan2013}. Na\"{i}vely, this should result in a similar anisotropic crystal field environment for the Eu moments in both LSAT epitaxial samples and bulk samples. This should render Eu moments in films to also orient in the $ab$-plane at zero field. Corroborating this, the $ab$-plane is the magnetic easy plane as probed by both film magnetoresistance and  magnetometry \cite{Ahadi2019, Takahashi2009}. Our earlier neutron diffraction measurements established that, at zero field, the G-type AFM order of the parent system persists in metallic samples ($n<9{\times}10^{20}$ cm$^{-3}$) with only a slightly decreasing onset temperature \cite{Ahadi2018}.
 
In this study, we report the evolution of magnetic order under an applied in-plane magnetic field in a series of Eu$_{1-x}$Sm$_x$TiO$_3$ samples with chemical potentials that span across the proposed Weyl nodes in this compound. Specifically, we examined an undoped insulating parent sample with $x=0$, a semimetallic sample with Fermi level very near the proposed Weyl node with $x=0.02$ ($n=4.0 \times 10^{20}$ cm$^{-3}$), and a fully metallic sample with $x=0.04$ ($n=7.8 \times 10^{20}$ cm$^{-3}$). Using a combination of neutron diffraction and polarized neutron reflectometry (PNR), we determine the average and depth-dependent magnetic structures. We find that, independent of doping, the application of external magnetic fields above $0.3$ T begins to quench AFM order and drive a spin-flop transition. Neutron reflectometry measurements on the sample close to the reported Weyl state reveal the presence of an interfacial layer where a FM state is stabilized at the substrate interface at fields as small as $0.5$ T. The implications of our findings for recent magnetotransport measurements are discussed.

\section{Methods}

Neutron scattering measurements were performed on 100 nm thick Eu$_{1-x}$Sm$_x$TiO$_3$ films grown epitaxially on (LaAlO$_3$)$_{0.3}$(Sr$_2$TaAlO$_6$)$_{0.7}$ (LSAT) (001) single crystal substrates using hybrid molecular beam epitaxy (MBE), the details of which have been reported elsewhere \cite{Ahadi2017}.  Neutron diffraction measurements were collected on the BT-7 triple-axis spectrometer at the NIST Center for Neutron Research \cite{Lynn2012}. Measurements were taken with initial and final neutron energies fixed at $E_i=E_f=14.7$ meV using a pyrolitic graphite (PG) monochromator and analyzer. One PG filter was placed before the sample and two after the sample to greatly reduce higher harmonic contamination. The collimator configuration was open-50$^\prime$-50$^\prime$-120$^\prime$ before the monochromator, sample, analyzer, and point detector, respectively. Samples were oriented in the [$H$ 0 $L$] scattering plane and mounted within a $^3$He insert in a 7 T vertical field continuous magnet cryostat. Film peaks are indexed to a tetragonal unit cell (space group $I4/mcm$, $a=5.48$ \AA, $c=7.9$ \AA$\;$for the film) throughout this paper. We also index the LSAT (001) substrate peaks to a tetragonal cell, due to the presence of anti-phase boundary reflections. Uncertainties represent one standard deviation in the data.

Polarized neutron reflectometry (PNR) measurements were performed using the PBR reflectometer at the NIST Center for Neutron Research with an incident wavelength of 4.75 \AA. The sample was mounted in the same sample environment used at BT-7, with the film's surface normal to the scattering wavevector $q$. PNR measurements were collected on field cooling at 2 T from 35 K. Reflectometry datasets were reduced \cite{Maranville2018} and refined to slab layer models using the Refl1D code, which implements an optical matrix formalism \cite{Chatterji2006, Kirby2012}.

\section{Experimental Results}

\begin{figure}
\subfigure{
\includegraphics[trim=0mm 0mm 0mm 0mm, clip,width=0.45\textwidth]{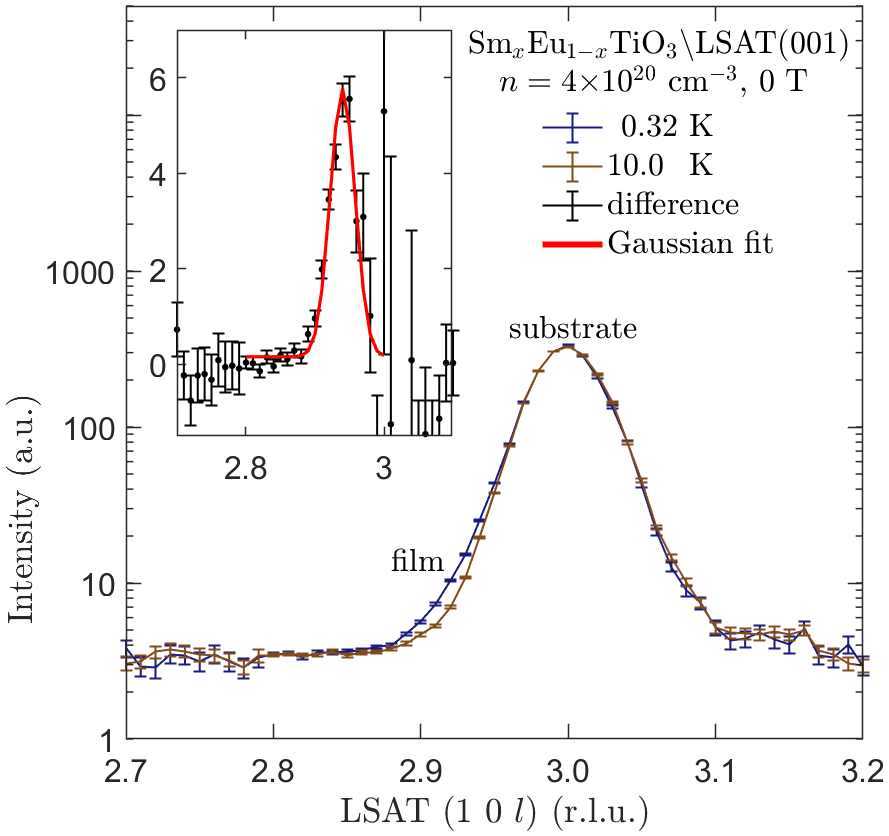}
\label{fig:Lpeak}
}
\caption{Magnetic neutron diffraction: raw diffraction scans along $l$ are shown both at base temperature (blue) and at the 10 K `background' (gold). Inset: the background-subtracted scan (black) reveals the G-type AFM Sm$_x$Eu$_{1-x}$TiO$_3$ peak. This is fit to a simple Gaussian (red) with fixed width and fixed center to extract the area. Widths were fixed to the instrumental resolution; in the case of (103), 0.018 r.l.u. along $l$ (7 nm = 9$c$).} \label{fig:peaks}
\end{figure}

We first discuss the results of neutron diffraction measurements probing the evolution of AFM in Eu$_{1-x}$Sm$_x$TiO$_3$ samples as a function of electron doping and in-plane magnetic field.  Given the presence of half-order reflections from the LSAT substrate at the film's AFM reflection conditions, care must be taken to resolve magnetic signal from the film against the large background of the neighboring substrate peaks. Due to a combination of the large Eu$^{2+}$ moment and elongation of the EuTiO$_3$ $c$ axis, this can be successfully achieved via analysis of magnetic Bragg peaks at tetragonal $[odd \;0\; odd]$ positions (here, the $(1 0 3)$, $(1 0 5)$, and $(3 0 3)$ in all three samples). As an example, the $(1 0 3)$ reflection of the intermediate $x=0.02$ doping sample is shown in Fig. \ref{fig:peaks}. The film is apparent as increased intensity along the low-$q$ side of the substrate peak and can be resolved via subtraction of the high temperature background as shown in the Fig. \ref{fig:peaks} inset. All the AFM reflections resolved in this way were instrument resolution-limited along the $h$ and $l$ axes, consistent with long-range G-type AFM order.

\begin{figure}
\begin{flushleft}
\subfigure{
\includegraphics[trim=0mm 0mm 0mm 0mm, clip,width=0.355\textwidth]{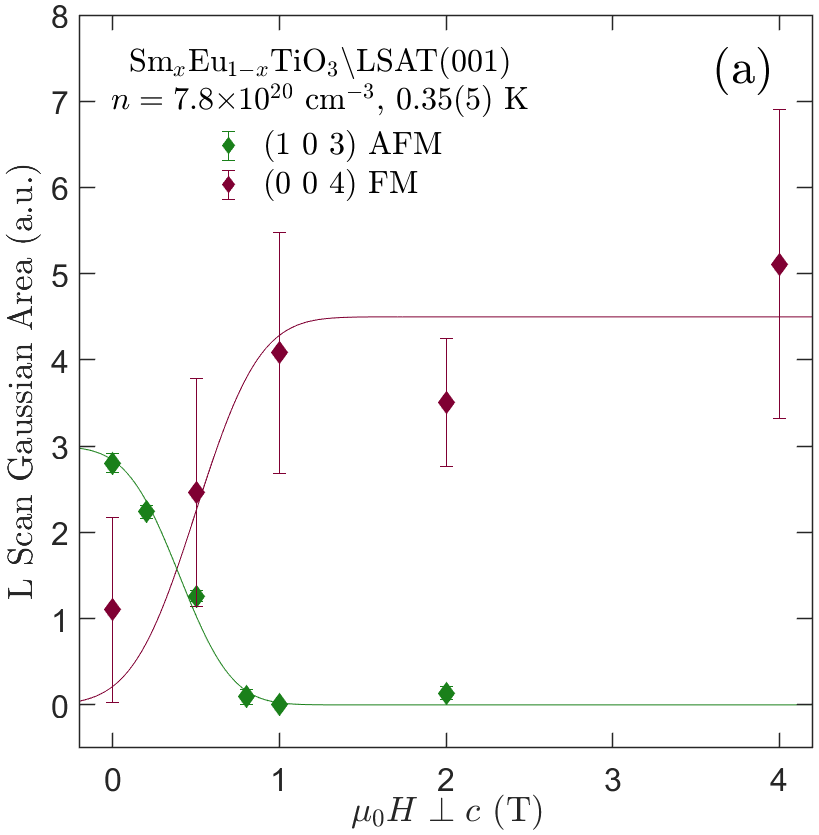}
\label{fig:metalOP}
}
\subfigure{
\includegraphics[trim=0mm 0mm 0mm 0mm, clip,width=0.405\textwidth]{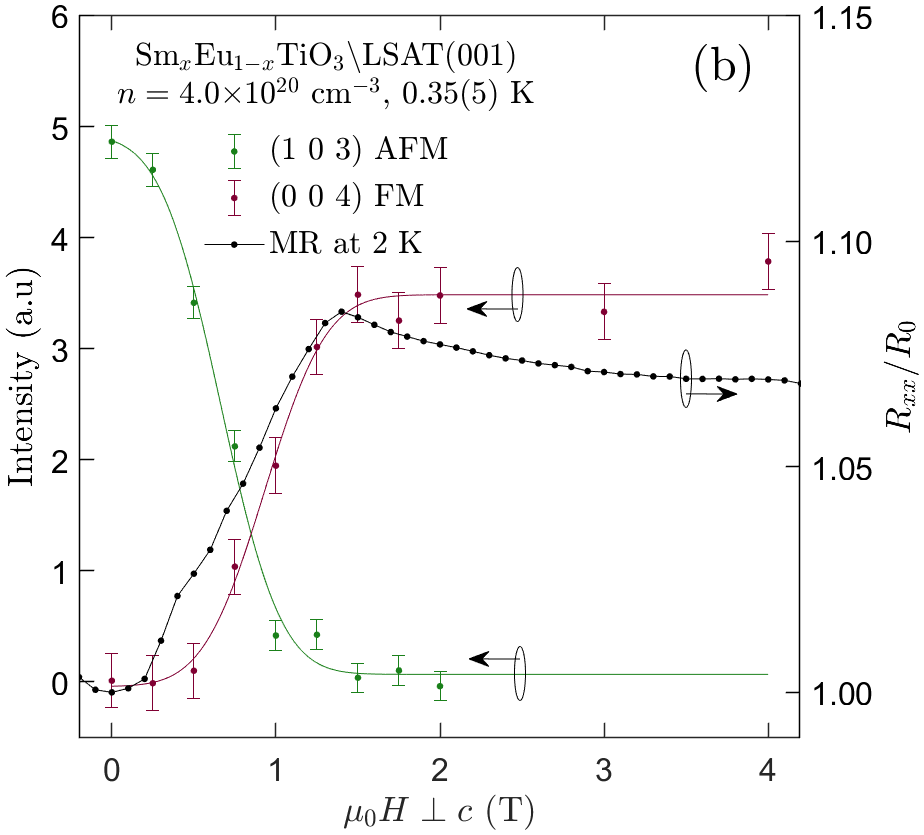}
\label{fig:intermediateOP}
}
\subfigure{
\includegraphics[trim=0mm 0mm 0mm 0mm, clip,width=0.355\textwidth]{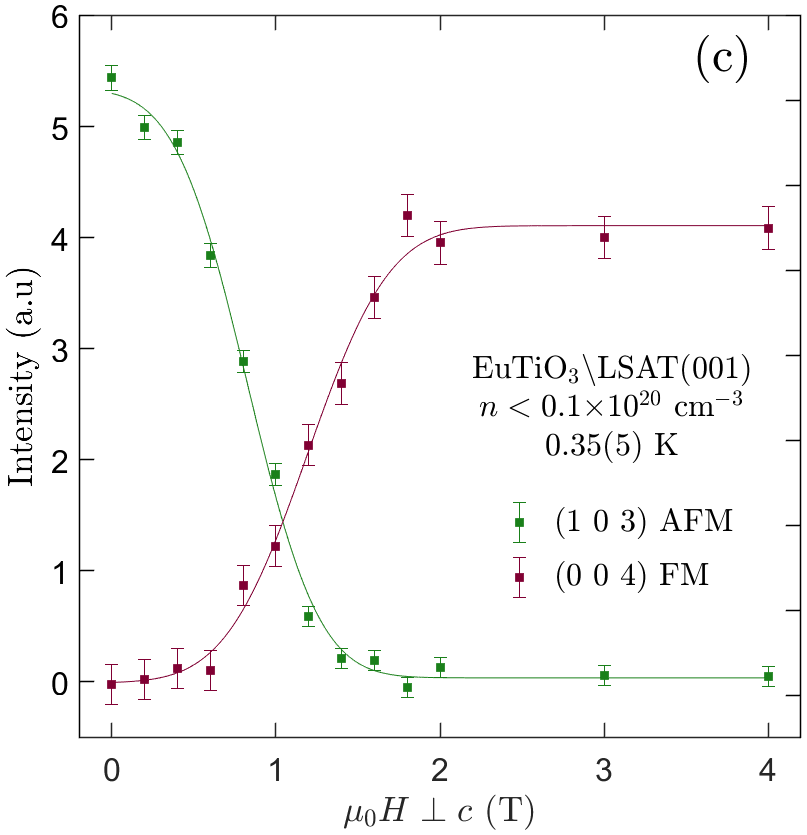}
\label{fig:parentOP}
}
\end{flushleft}
\caption{Order Parameters: AFM peaks in green and FM peaks in red are shown for (a) metallic (b) semimetallic and (c) parent Sm$_x$Eu$_{1-x}$TiO$_3\backslash$LSAT(001) samples. Data are shown in symbols, and lines are error function fits to the data. In (b), the black symbols are magnetoresistance taken at 2 K with $I || H{\perp}c$ for a similarly doped sample $n=3.4 \times 10^{20}$ cm$^{-3}$ which is 50 nm thick. Diffraction samples are 100 nm thick. Details are in the text.} \label{fig:OPs}
\end{figure}

With applied field, we observe a reflection corresponding to a partial ferromagnetic polarization of the Eu moments at the $(0 0 4)$ wave vector. Upon increasing field, the intensity of this peak and related ferromagnetic reflections is enhanced, and the AFM peaks diminish. This is indicative of the sample undergoing a continuous spin-flop transition.  Looking first at the metallic, highly doped $x=0.04$ sample, the aforementioned method of measuring $l$-scans was used to track both the AFM $(1 0 3)$ and FM $(0 0 4)$ components of the canted magnetic order under increasing field strengths through the spin-flop transition. The peak areas from fits to $l$ scans through both reflections are plotted as a function of field in Fig. \ref{fig:metalOP}. 

After we measured this transition in the metallic $x=0.04$ sample, the spin-flop transitions in the other two $x=0$ and $x=0.02$ samples were parameterized by first confirming the position of the $(1 0 3)$ and $(0 0 4)$ peaks with $l$-scans at zero-field and high-field, respectively, followed by counting at the peak positions upon sweeping field. These results are shown in Fig. \ref{fig:intermediateOP} and \ref{fig:parentOP} for the $x=0.02$ and $x=0$ samples, respectively, and the difference in measurement methodology is the reason for difference in the relative sizes of the error bars.

The field-driven canting of the zero-field AFM order, summarized in Fig. \ref{fig:OPs}, shows a suppression of the G-type AFM reflection intensity with increasing field for all samples that occurs in a manner largely independent of doping level or metallicity. 
The nominal spin-flop transition begins at fields as small as 0.25 T with a small but statistically significant intensity decrease in all but the intermediate doped sample, for which the intensity decrease is within uncertainty. Most of the AFM signal suppression occurs between 0.25 T and 1.5 T, above which, the Eu moments are fully polarized in all samples. This behavior is consistent with prior studies of the bulk magnetization \cite{Petrovic2013}. Analysis of the full width at half maximum (FWHM) along $l$ and $h$ at representative fields shows that the AFM peaks remain resolution-limited throughout the entire range of fields, indicating that three-dimensional long-range order is maintained throughout the transition. 

\begin{figure}
\includegraphics[trim=0mm 0mm 0mm 1mm, clip,width=0.48\textwidth]{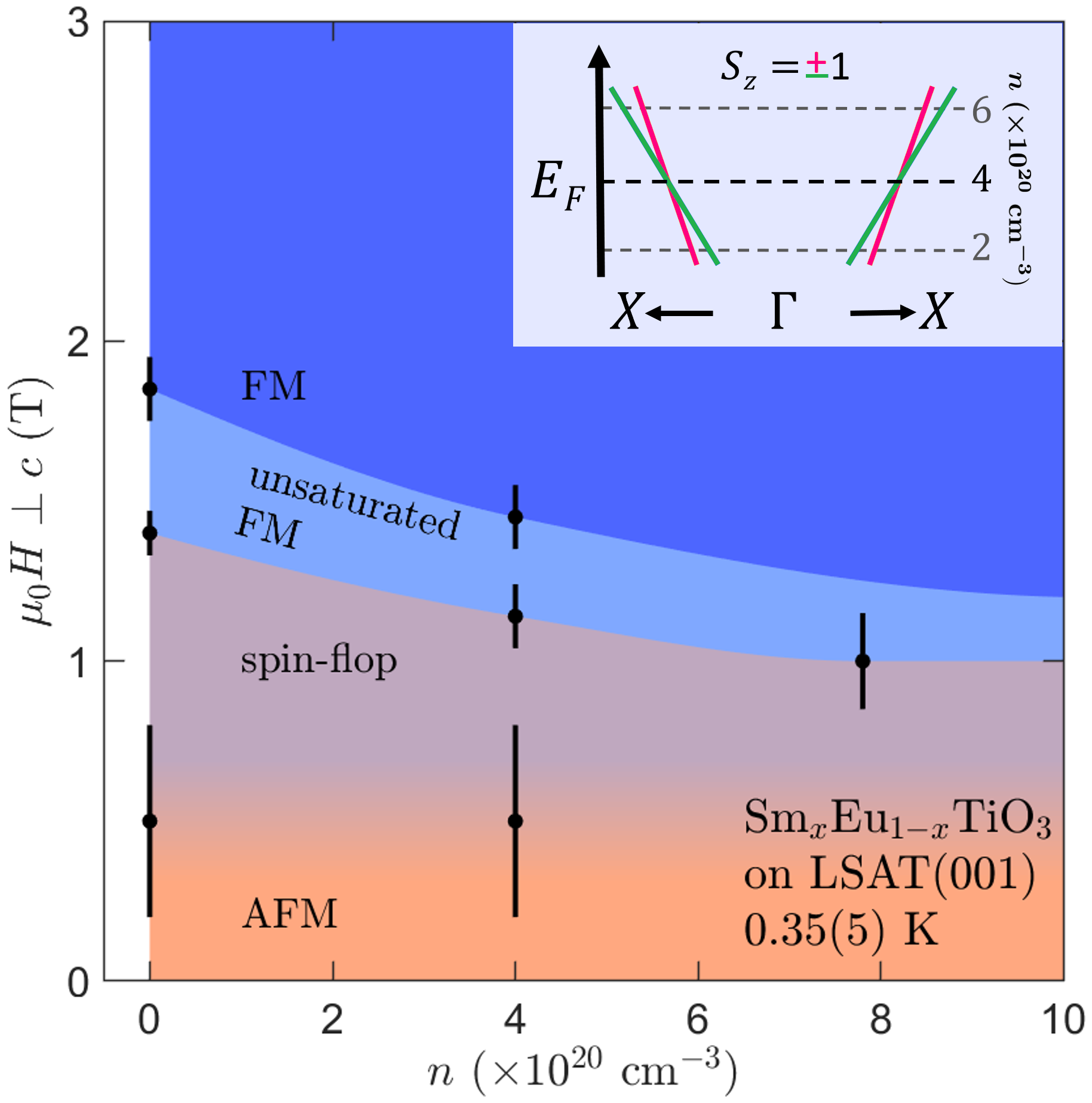}
\caption{ Field-Concentration phase diagram based on diffraction measurements of the parent, intermediate, and metallic samples. Label abbreviations are `AFM' antiferromagnetism and `FM' saturated ferromagnetism. The applied field is in-plane, perpendicular to the tetragonal $c$ axis. Error bars are from fit uncertainties in Fig. \ref{fig:OPs}. Colored regions are speculative guides to the eye. Inset: Sketch of one predicted pair of conduction band crossings, projected on the spin $z$-component. Horizontal dashed lines indicate the Fermi energy at different carrier concentrations. DFT calculations predict Weyl nodes for $n\approx4{\times}10^{20}$ cm$^{-3}$ near ($\pm 0.24 \pi/a$, 0, 0). Adapted from Ref. [\onlinecite{Ahadi2018}].} \label{fig:Bn}
\end{figure}

The field dependence for both the FM and AFM components of the magnetic order can be parameterized via error functions, which allow a systematic extraction and analysis of saturation and crossover fields. We used error functions of the form $\frac{A}{2}[1{\pm}erf(\frac{\mu_0 H-\mu}{\sqrt{2}\sigma})]$ for the FM (AFM) components, with free parameters magnitude $A$, center $\mu$, and variance $\sigma$. Critical fields $B_{on}$ ($B_{off}$) were defined as 5$\%$, and the saturation field $B_{sat}$ to be 95$\%$, of the maximum fit value for FM (AFM) order. These values are used to define the phase boundaries in Fig. \ref{fig:Bn}.

Compiling our diffraction data, we depict a magnetic phase diagram as a function of carrier concentration. We highlight the AFM phase, the spin-flop transition, and the unsaturated and saturated FM phases. The thresholds for field-polarization decrease with electron concentration, consistent with conduction electrons destabilizing the AFM ground state. We note that field-polarized FM order is saturated below 3(1) K yet persists to 13(1) K at 3 T for the metallic sample, in agreement with features in the magnetoresistance \footnote{See Supplemental Material at [URL will be inserted by publisher] for the ferromagnetic order parameter and polarized neutron reflectometry refinements with different magnetization maximum bounds.\label{note1}}.

\begin{figure}
\subfigure{
\includegraphics[trim=0mm 0mm 0mm 0mm, clip,width=0.5\textwidth]{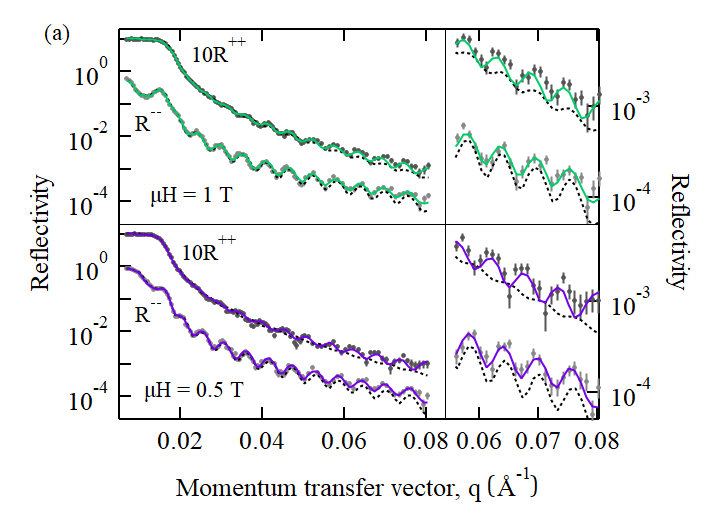}
\label{fig:PNR_data}
}
\subfigure{
\includegraphics[trim=0mm 0mm 0mm 0mm, clip,width=0.45\textwidth]{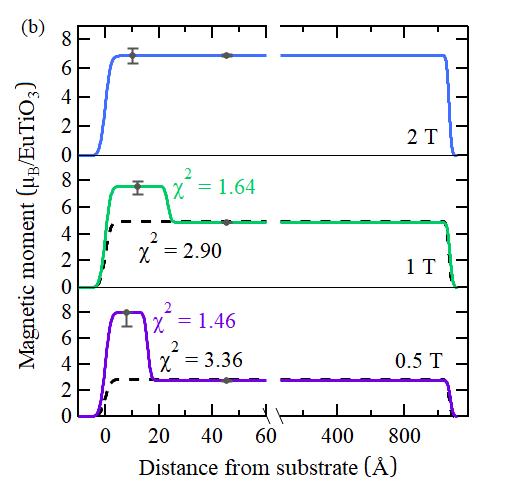}
\label{fig:PNR_SLD}
}
\caption{ Polarized Neutron Reflectometry: (a) data for the semimetallic intermediate doping sample ($n=4.0 \times 10^{20}$ cm$^{-3}$) at various applied fields at 0.35 K along with refined fits for magnetically uniform (dashed line) and interfacial magnetism (solid lines) models. (b) Magnetic scattering length density models from which the fits to the data were calculated. Grey data points are the best fit refined magnetization value for each slab in the two slab model with error plotted as the 95\% confidence interval.} \label{fig:PNR}
\end{figure}

The diffraction measurements described report on the volume-averaged field evolution of magnetism in the Eu$_{1-x}$Sm$_x$TiO$_3$ sample series. However, it is well known that the properties of thin films may deviate from the bulk behavior due to effects such as strain or charge transfer near heterointerfaces \cite{hellman2017interface}. Therefore, we utilized PNR to further investigate the intermediate $x=0.02$ doped sample's depth-dependent, magnetization profile. In doing so, we observed a magnetic interfacial layer adjacent to the substrate, where the saturated FM state is stabilized at lower fields than seen in our bulk-sensitive diffraction measurements. The following paragraphs describe how this interfacial layer was resolved.

The magnetic depth profile was determined by fitting two comparative models to the measured and reduced reflectivity data, shown for lower fields (i.e. 0.5 T and 1 T) in Fig. \ref{fig:PNR_data}. The first model considered was a uniformly magnetized film, and the corresponding best fit to the data is shown by the black dashed curves in Fig. \ref{fig:PNR_data}. As seen in the righthand panels, which highlight the high $q$ region, a uniform magnetization model does not capture the falloff of the reflectivity curve properly. Specifically looking at the R$^{++}$ scattering channel, it can be seen that a uniform magnetization model results in a flattened oscillation amplitude that is much smaller than the measured data. 

Therefore, we considered a second model with two magnetic slabs to simulate a non-uniform magnetic depth profile. Each slab in the model was free to fit both its thickness and magnetization. The best fit results of this model are shown by the solid color curves in Fig. \ref{fig:PNR_data} and correspond to the $\chi^2$ values in colored font. Clearly, these non-uniform models capture the high $q$ oscillations missed by the uniform magnetic model. The improved fit quality can also be seen in the numerical goodness of fit for each model. It is important to note that, in testing the two slab magnetization model, the refinement was initialized in a variety of starting conditions (i.e. thin layer at substrate, thin layer at surface, equal layer thicknesses) and that regardless of how the model was initialized, the end result converged to an interfacial layer near the substrate and not at the film's outer surface. 

The magnetization models corresponding to these best two-slab fits are plotted in Fig. \ref{fig:PNR_SLD} in units of moment per Eu-ion and color coded to match the reflectivity fits in Fig. \ref{fig:PNR_data}. First, at 2 T, the depth profile is that of a fully saturated film with 6.92(4) $\mu_{B}$/Eu, very near the expected 7 $\mu_{B}$/Eu expected for fully polarized S = 7/2 Eu moments. Importantly, here the model ``chooses" to be magnetically uniform even when modeled as two independent magnetic slabs, which is consistent with neutron diffraction results suggesting that a 2 T field is sufficient to saturate the system. However, when the field is lowered to 1 T, the majority of the sample volume has a reduced magnetization of 4.88(3) $\mu_{B}$/Eu following the trend expected from our diffraction results. Yet near the interface, we observe a region approximately 2 nm in thickness that remains in a saturated state. The enhanced magnetism in this interfacial layer persists to fields as low as 0.5 T, the lowest field at which we collected PNR data, even as the bulk of the film continues to reduce its net magnetization. 

In the two-slab PNR refinements discussed above, the magnetization in each slab was allowed to vary up to 8 $\mu_{B}$/fu to account for the possibility of a Ti 3d$^1$ moment co-aligned with the Eu moments (which would only be germane to the interface layer). This procedure resulted in refined interface magnetization values of 7.6 and 8 $\mu_{B}$/fu for the 1 T and 0.5 T data sets, respectively. However, crucially, the error on this interfacial magnetization is large (95\% confidence interval $\approx$ 6.9 to 8 $\mu_{B}$/fu) and overlaps with both the refined 2T value and the theoretical S = 7/2 value. In other words, we cannot distinguish between models with and without a moment on the Ti site. To prove this point, in the Supplemental Materials we present the results of refinements in which the interfacial moment is given an upper bound of 7 $\mu_{B}$/fu. The results and conclusions are essentially identical to those in the main text, with changes in $\chi^2$ and magnetization in the thick, upper EuTiO$_3$ layer being smaller than the error. Only the interfacial layer thickness changes notably, increasing an average of 18\% to offset the 12\% reduction in magnetization. Further analysis of the covariance matrix shows these two parameters are highly correlated can be traded for one another in the modeling. Despite this coupling of parameters, our data nevertheless demonstrate that the interface layer remains robustly FM in the low field regime where the bulk of the sample behaves as a canted antiferromagnet.

\section{Discussion}

Our neutron diffraction results and recently published anisotropic magnetoresistance (AMR) measurements paint a coherent picture of the magnetic structure evolution as a function of field. EuTiO$_3$ films with or without Sm doping are G-type antiferromagnets at zero field. This AFM state remains relatively stable until approximately 0.3 T as shown in Fig. \ref{fig:OPs}, at which point we observe the onset of a FM signal coincident with the initial reduction of AFM order. This field strength is, within error, the same field at which bulk samples of the parent EuTiO$_3$ undergo a spin-flop transition and the Eu moments reorient from the $c$-axis into the $ab$-plane \cite{Petrovic2013}. Moreover, 0.3 T is where the onset of the AMR signal occurs in doped thin film samples \cite{Ahadi2019}. Demonstrating this, in Fig. \ref{fig:intermediateOP}, the longitudinal magnetoresistance (MR) of an intermediately electron doped sample is plotted (solid black dots) alongside the diffraction data and shows remarkable similarity to the development of the FM component of magnetic order. Together these data suggest that 100 nm thick EuTiO$_3$ films, independent of doping, are thick enough that they retain a bulk-like spin-flop transition near 0.3 T in which the moments reorient into the plane of the film and retain the same propagation vector.

Further connecting to prior AMR measurements, Fig. \ref{fig:OPs} and Fig. \ref{fig:Bn} show the spin-flop transition proceeds smoothly until the AFM component of the ordered state is completely suppressed between 1 T and 1.5 T, depending on the doping level. This is remarkable since, at 0.8 T, AMR data shows a distinct transition wherein the four-fold symmetric AMR signal undergoes a 45$^{\circ}$ rotation (i.e. a rotation of the AMR maximum from [1 0 0] to [1 1 0]).  K. Ahadi, \textit{et al.} attribute this transition to changes in the electronic structure, such as the movement of Weyl nodes \cite{Ahadi2019}. The absence of any abrupt features in our diffraction data near 0.8 T supports this hypothesis, though there is considerable measurement uncertainty which could obscure subtle magnetic transitions ${<}$ 1 $\mu_B$ over a narrow range of magnetic field. Our data are instead consistent with a continuous canting of the moments within the $ab$-plane, from antiparallel at 0.3 T to parallel above 1 T. Within this picture, we can then use the absolute Eu moment determined from the bulk slab in our PNR refinements to determine the evolution of canting as a function of field for the intermediate doped sample. Specifically, we find the canting angle to be 23.4(2)$^{\circ}$ and 44.2(2)$^{\circ}$ at 0.5 T and 1 T fields, respectively. 

Between 1.3 and 2 T, FM order saturates. This agrees well with a final transition in AMR measurements where the symmetry gradually reduces from four-fold to two-fold over this same field range \cite{Ahadi2019}. This two-fold symmetry, hysteresis in the anomalous Hall signal, and additional intensity on the zone center diffraction peaks all indicate that the high-field state is one of field-induced FM order. Importantly, both the FM saturation field and the AFM order suppression field decrease with increased doping; see Fig. \ref{fig:Bn}. This reveals that the addition of itinerant electrons to the system destabilizes the AFM state, in agreement with recent theoretical results \cite{Gui2019}. Similar behavior is seen in doped bulk crystals, where magnetization and electron transport studies on very highly electron-doped bulk crystalline samples (near 0.1 additional electrons per Ti) have previously shown that AFM order is quenched and an unusual field-polarized ferromagnetic (FM) order becomes dominant \cite{Katsufuji1999}.

These results combined with our observation of interfacial FM stabilized at the substrate interface are consistent with a picture of magnetism in EuTiO$_3$ films wherein AFM and FM ground states are energetically close and slight changes in orbital filling can tip the balance \cite{Akamatsu2011}. There exist several microscopic theories to explain the propensity for FM order, including: an indirect Eu $5d$ exchange \cite{Akamatsu2011}, an RKKY interaction between Eu $4f$ and Ti $3d$ states \cite{Takahashi2009}, or coupling between Eu $5d$ and Ti $3d$ states in the conduction band \cite{Gui2019}. However, additional work is needed to determine which of these explanations, or which combination of them, is operational within different doping and strain regimes. In this regard, the observation of the FM interfacial layer in our PNR data provides an intriguing experimental geometry in which to probe and decouple these various possible stabilizing contributions, and follow-up studies along these lines are in progress.

In prior transverse magnetotransport measurements, there is a distinct change in the sign of the AHE as a function of doping \cite{Ahadi2018}. However, our neutron measurements show little difference in the magnetic order under applied field between differently doped samples with the exception of a slight change in the FM saturation values and threshold spin-flop fields discussed above. Rather, the sign of the AHE resistance changes with doping in a manner consistent with the chemical potential crossing a Weyl node, as sketched schematically in the inset to Figure \ref{fig:Bn}. For Weyl semimetals, the AHE is known to be proportional to the net magnetization \cite{Takahashi2018}, with the magnitude dependent to simplistic approximation on the Weyl node separation in reciprocal space \cite{Armitage2018}. In the present case, the AHE magnitudes are uncorrelated with the diffraction results---suggesting that the electronic state is weakly informed by the average Eu magnetism.  We also propose that the interfacial FM layer may itself be responsible for some unusual magnetotransport signatures in this system. The hypothesis that magnetic inhomogeneity can potentially `mimic' a topological Hall effect, as reported for SrRuO$_3$ \cite{Kan2018}, is an avenue that should be explored.

\section{Summary}

In summary, we report neutron scattering measurements of the evolution of long-range magnetic order as a function of both magnetic field and electron doping in the Weyl semimetal candidate Eu$_{1-x}$Sm$_x$TiO$_3$ on LSAT (001). The evolution of magnetic order under an applied, in-plane magnetic field correlates well with previous magnetrotransport data reporting unconventional AMR effects.  The persistence of an AFM ground state and similar spin-flop field thresholds as a function of electron doping suggest the presence of an additional electronic phase transition at intermediate fields in semimetallic $x=0.02$ samples, as well as a nonmagnetic origin of the apparent topological Hall effect.  Using polarized neutron reflectometry to probe depth-dependent magnetism in the semimetallic sample, we resolve a saturated ferromagnetic interfacial layer that forms at the substrate interface in doped samples prior to the completion of the spin-flop transition throughout the bulk of the film.  The impact of this inhomogeneity within the magnetic depth profile in transverse magnetotransport measurements warrants future study.

\begin{acknowledgments}
This work was supported by NSF award DMR-1729489 (S.D.W. and Z.P.). R.F.N. acknowledges support from the National Research Council Research Associateship Program.
\end{acknowledgments}

\bibliography{eutio3_ND_edits}

\end{document}